# Pan-African Asteroid Search Campaign: Africa's Contribution to Planetary Defense

Miracle Chibuzor Marcel[1], Kassamba Abdel Aziz Diaby[2], Meryem Guennoun[3], Betty Rose Nabifo[4], Mohamed Elattar[5], Andoniaina Rajaonarivelo[6], Privatus Pius[7], Molly Nkamogelang Kgobathe[8], Immanuel Luis[9], Sigrid Shilunga[10], Nejmeddine Etteyeb[11], Keketso Qhomane[12], Samuel Nyangi[13], Tresford Chilufya Kalunga[14], Nunes Alfredo Assano[15], Edson Domingos Jequecene[16], MAFUKA LUSALA JOSEPH[17], Esaenwi Sudum[18], Jorbedom Leelabari Gerald[19], Christopher Tombe Louis Gore[20], Kareem Waleed Hosny[21], Nagat Yasser[22], Jocelyn Franck[23], MAMOUDOU KOUROUMA[24], BABOUCARR BOBB[25], Kebab Jaiteh[26], Salma Sylla[27], Hans ESSONE OBAME[28], Dennis Kiyeng[29], Thobekile Sandra Ngwanw[30], Tawanda Kelvin Simon[31], Saja Alhoush Sulayman[32], Salma Regaibi[33], Souley Yahaya[34], Tengwi Mogou Ornela[35], Henry Sanderson Viyuyi[36], Fortune Tatenda Matambo[37], Matthias Asare-Darko[38], Christian Kontoa Koussouwa Gbaba[39], Moisés da Silva[40], Ntahompagaze Joseph[41], Gilberto Gomes[42], Bongiwe Portia Mkhabela[43], Bauleni BVUMBWE[44], Tshombe Nkhowani[45], Mawugnon Axel Gahou[46], Sarah Abotsi-Masters[47], René Simbizi[48], Salomon Mugisha[49], Ahmed Saeed[50], Mohammed Yahya Alradi Eldaw[51], Allen Thomas[52], Ben Abdallah ridha[53], Dieumerci kaseha[54], Sherine Ahmed El Baradei[55], Nahla Hazem Hussein[56], BADO Fabrice[57], Ngozika Frances Anekwe[58], Arvind Ramessur[59], Mohamed Ali Koroma[60], Harold Safary[61], Oosthuizen Leonardo[62], Mdumiseni Wisdom Dabulizwe Dlamini[63], Mamadou Mahamat Djabbi[64], Nonofo Angela[65], Mamaja Jalloh[66], Mamadou Balde[67], Joy Olayiwola[68], Elijah Ibharalu[69], Thierry Martial TCHANGOLE[70], Kirubel Memberu[71], Lidia Dinsa[72], Chidozie Gospel Ezeakunne[73]

1. Pan-African Citizen Science e-Lab, FCT, Abuja: info@pacselab.space, miracle.c.marcel@gmail.com; 2. Université Félix Houphouët-Boigny; 3. Oukaimeden Observatory; 4. National Curriculum Development Centre; 5. IEEE New Beni Suef; 6. Haikintana Astronomy Association; 7. Department of Natural Science, Mbeya University of Science and Technology; 8. University of Botswana Astronomy Club; 9 & 10. Department of Physics, Chemistry and Material Science University of Namibia; 11. Tunisian Astronomical Society; 12. BlueCraneSpace Astronomy & Astrophysics Department - University of Pretoria; 13. Amateur Astronomical Society of Kenya; 14. Kabulonga Girls' Secondary School; 15 & 16. Detetives do Cosmos; 17. Astroclub Kongo Central; 18. The Astro Group of the Rivers State University; 19. The Astro Group of the Rivers State University; 20. Mayardit Academy for Space Sciences - University of Juba; 21 & 22. Pharaohs of Space; 23. University Marien Ngouabi; 24. IT Dreams and Promises of the University of Bamgui; 25. Physics Dept. University of Gambia; 26. Physics Dept. University of Gambia; 27. Orion Astrolab and Department of Physics, University Cheikh Anta Diop; 28. IAU - NAEC Gabon; 29. Space Partnerships and Applications Company Kenya; 30. Zimbabwe Astronomical Society; 31. Zimbabwe National Geospatial and Space Agency; 32. Amateur Astronomy Libya; 33. Steps Into Space Association; 34. Niger Space Surfer; 35. Astronomy Club of the University of Buea; 36. Zambia Space Explorers; 37. Zimbabwe National Geospatial and Space Agency; 38. PRAGSAC; 39. NGO Science Géologique pour un Développement Durable (SG2D); 40. Associação Angolana de Astronomia; 41. Physics Department, University of Rwanda, College of Science and Technology; 42. Angolan Space Program Management Office (GGPEN); 43. Galaxy Explorers; 44. Celestial Explorers; 45. Copperbelt University; 46. Sirius Astro-Club Benin; 47. Ghana Planetarium; 48 & 49. Physics Dept. University of Burundi; 50. Sudanese Asteroid Hunters; 51. Institute of Space Research and Aerospace (ISRA); 52. Center for Science Education; 53. Aljarid Astronomie; 54. Lubumbashi Astro Club; 55 & 56. Space -Water-Environment Nexus e - Center; 57. Laboratoire de Physique et de Chimie de l'Environnement, Université Joseph KI-ZERBO, Ouagadougou, Burkina Faso; 58. Phys. Dept. Chukwuemeka Odumegwu Ojukwu University Uli Campus Anambra State; 59. IAU - NOC Mauritius; 60. Cosmic Gazers Research Institute Sierra Leone; 61. Kenya Space Agency - Education and Outreach Students' Network; 62. Night Sky Tours; 63. Destiny Stars of the Univesity of Eswatini; 64. Toumaï; 65. Marang Junior Secondary School; 66. Sierra Leone Geospatial and Space Agency; 67. UNESCO Center for Peace USA in Guinea; 68. National Space Research and Development Agency, NASRDA HQ; 69. University of Benin; 70. CosmoLAB Hub Association; 71 & 72. Ethiopian Space Science Society; 73. University of Central Florida

**Abstract**

Asteroid search is a global effort for planetary defense. The International Astronomical Search Collaboration (IASC) is the leading global educational outreach program that provides high-quality astronomical datasets to citizen scientists to discover asteroids. Since December 4, 2020, the Pan-African Citizen Science e-Lab (PACS e-Lab) has been IASC's biggest partner on the continent in recruiting and training citizen scientists in asteroid search endeavors. Over 30 asteroids have been discovered by 60 citizen scientists. About 595 citizen scientists from over 40 countries have been engaged in the project up to the time of composing this literature. The group is set to expand its endeavors to the rest of the continent and aims to engage thousands of citizen scientists.

## 1.0 Introduction

Asteroids are remnants from the early solar system that could not form into a definite celestial body, like planets. Most of them settled between the orbits of Mars and Jupiter, in a region called the Main Belt (Marchi et al., 2022). However, events like gravitational perturbations from Jupiter (Marcos & Marcos, 2016), asteroid-to-asteroid collisions (Morbidelli et al., 2015), the Yarkovsky effect (Elía & Brunini, 2007), etc., can affect Main Belt asteroids by causing them to migrate either toward the inner orbit or outer orbit. These asteroids with distorted orbits have names like Mars-crossing asteroids (MCA) (Hou, 2024), Near-Earth Asteroids (NEA) (Hou, 2024), Trans-Neptunian Objects (TNOs) (Fernández-Valenzuela et al., 2015), Jupiter-Crossing Asteroids (JCA) (Morbidelli et al., 2005), Mercury-Crossing Asteroids (MCA) (Gladman & Coffey, 2009), Venus-Crossing Asteroids (VCA) (Schaber et al., 1992), etc.

Near-Earth Asteroids are of particular interest in planetary defense because their close approach to Earth puts life forms in danger. This is because they sometimes fall to Earth and could wreak havoc (Bacu et al., 2023). One such remarkable asteroid impact event was on February 15, 2013, when a 20-meter diameter asteroid known as the Chelyabinsk meteor fell near Chelyabinsk in Russia. The impact of this asteroid generated a powerful shockwave that traveled across the nearby city, causing widespread damage to properties and injuring approximately 1,500 people (Попова et al., 2013).

On June 30 of the same year, the United Nations declared it World Asteroid Day to raise awareness on Planetary Defense against asteroids and the potential hazards they pose to our planet (Hucht, 2018). Consequently, there has been a rise in the number of surveys, both space-based and ground-based, set up to scan the sky (Mainzer et al., 2012). A few of these include the Panoramic Survey Telescope & Rapid Response System (Pan-STARRS) (Chambers et al., 2016) at the Institute for Astronomy at the University of Hawaii and the Catalina Sky Survey (CSS) (Larson et al., 2003) at the Lunar & Planetary Laboratory at the University of Arizona. These telescopes use automated pipelines to detect asteroids. However, sometimes these pipelines are not able to find asteroids due to reasons like fill factor effects, signal-to-noise requirements, and non-optimal sky-plane motion (Miller et al., 2024). The datasets taken from these surveys are sent to the International Astronomical Search Collaboration (IASC) so that, with the help of citizen scientists, the elusive asteroids can be found using the Astrometrica Windows OS program (Miller et al., 2024).

IASC was founded by Dr. Patrick Miller at Hardin-Simmons University in 2006 as an educational outreach program that provides high-quality astronomical datasets to the members of the public called citizen scientists who can contribute to the discovery and tracking of asteroids, comets, and other celestial objects (Miller, 2018). IASC works with several educational entities around the world that engage teachers, students, and space enthusiasts in the asteroid research exercise.

The Pan-African Citizen Science e-Lab (PACS e-Lab) was established to promote hands-on activities in astronomy & space science through citizen science and Soft Astronomy research in Africa as a means of advancing space exploration and enhancing space education and outreach. Starting from December 4, 2020, PACS e-Lab has been IASC's biggest partner in Africa (Marcel et al., 2024).

The objective of this study is to document and analyze the collaborative efforts between PACS e-Lab and IASC in promoting asteroid search across Africa.

## 2.0 Methodology

2.1 Collaboration Framework

The partnership between the Pan-African Citizen Science e-Lab (PACS e-Lab) and the International Astronomical Search Collaboration (IASC) was established to foster asteroid search activities across Africa, including North Africa and Sub-Saharan regions. The collaboration framework was designed to leverage the strengths of both organizations: PACS e-Lab connects with various educational entities across the continent and conducts training with them, while IASC provides asteroid datasets and analysis tools. The formal agreement between the two entities was facilitated by the African Astronomical Society through Dr. Charles Takalana.

2.2 Training and Outreach

Effective training and outreach were critical to the success of the asteroid search initiative. Groups were contacted via social media platforms like Facebook and LinkedIn, Google searches, emails, and referrals from individuals already involved in the project. Most of these groups responded positively, leading to introductory and training online meetings.

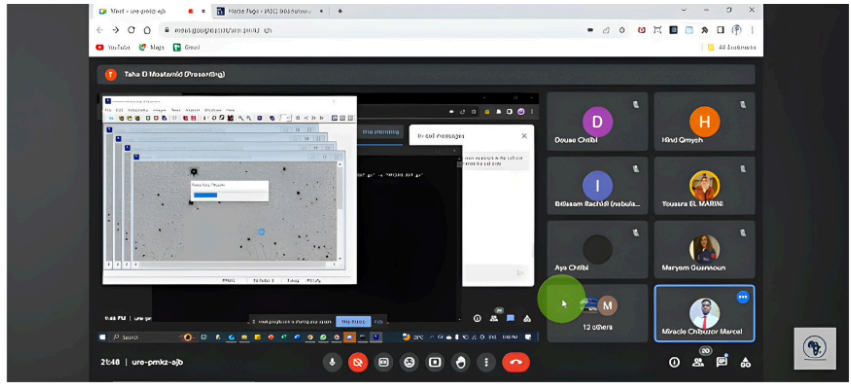

*Figure 1: One of many PACS e-Lab's online training sessions on asteroid search with teams across the continent, each lasting over three hours*

The online training sessions cover Astrometrica software installation, asteroid dataset analysis, preparing Minor Planet Center reports, reporting the results to IASC, and campaign registration. PACS e-Lab also offers recorded tutorials, which are particularly suitable for participants in locations where the internet is expensive or slow.

Regarding non-English speaking participants, PACS e-Lab has several volunteers from different locations who speak multiple languages. These volunteers are required to train their recruits and are sometimes called upon to train other participants speaking their language from another country or region. The volunteers have developed training materials, including manuals and video tutorials. These materials are available in multiple languages such as Arabic, French, and other local languages spoken in Africa to cater to the diverse linguistic landscape of the continent.

2.3 Data Collection and Analysis

Each participating team is made up of a minimum of two persons, with no specified maximum number. After the team has been trained in asteroid search, they are registered under IASC's Pan-African Asteroid Campaign to receive fresh datasets for analysis. Each dataset contains four FITS (Flexible Image Transport System) images taken at different time intervals of 30-60 minutes (Miller et al., 2024). These FITS images are taken and shipped to IASC from the Pan-STARRS and the Catalina Sky Survey. PACS e-Lab opted for the campaign to be run every month to ensure consistent hands-on activities. The citizen astronomers use Astrometrica software, designed for asteroid detection and measurement, to analyze the images and prepare a Minor Planet Center (MPC) report. The MPC reports of the analyzed images, including potential asteroid detections, are submitted to IASC for validation. IASC experts review the submissions and provide feedback, ensuring the scientific accuracy and integrity of the findings.

**Table 1**: PACS e-Lab's Calendar for the Asteroid Search for the 2023/2024 Academic Year

| 2023 | 2024 |
|---|---|
| August 10 – September 4 | January 5 - January 30 |
| September 8 – October 4 | February 5 - February 29 |
| October 9 – November 3 | March 6 - April 1 |
| November 7 - December 4 | April 5 - April 30 |
| Nil | May 3 - May 28 |

**Table 2**: PACS e-Lab's Calendar for the Asteroid Search for the 2024/2025 Academic Year

| 2024 | 2025 |
|---|---|
| August 1 – August 26 | January 23 - February 17 |
| August 29 - September 23 | February 21 - March 19 |
| September 26 - October 21 | March 24 - April 18 |
| October 25 - November 19 | April 23 - May 19 |
| November 22 - December 18 | May 23 - June 17 |

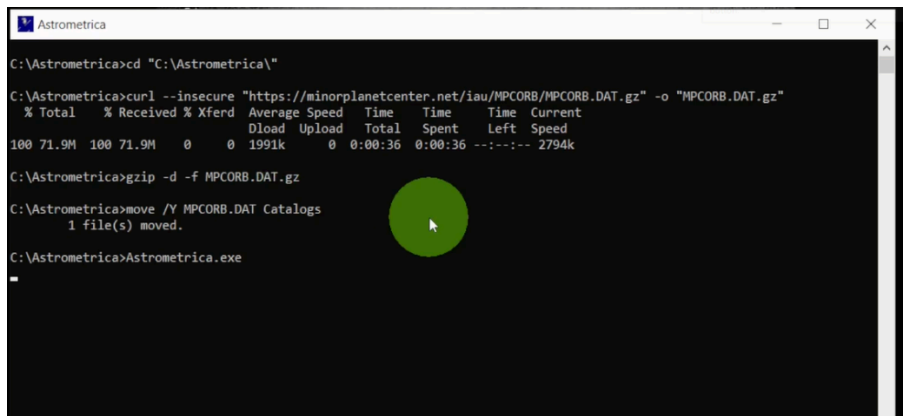

*Figure 2: Astrometrica Windows interface showing the loading of the Minor Planet Center*

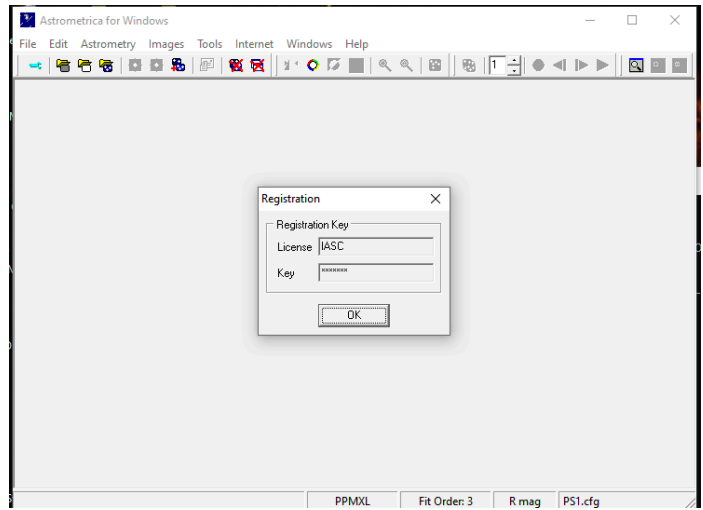

*Figure 3: Astrometrica Windows interface showing registration of the software with codes issued by IASC*

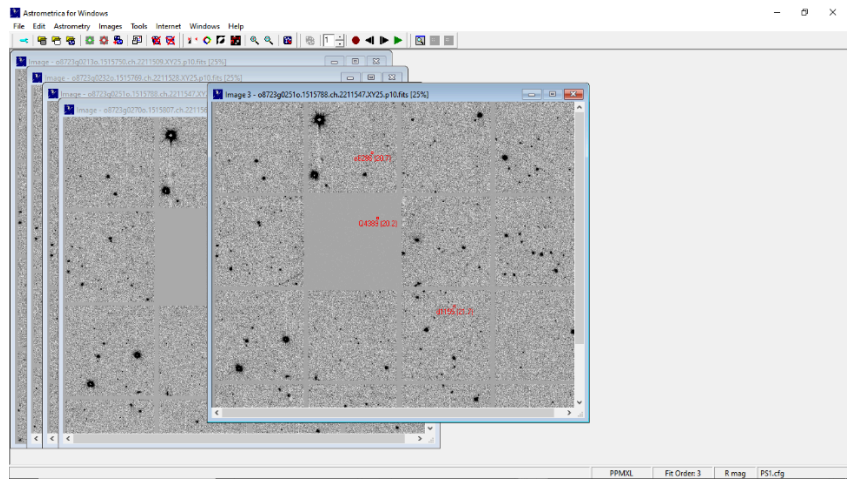

*Figure 4: Images (4) loaded and Blink function activated*

*Figure 5: Minor Planet Center Report for asteroid detection prepared with Astrometrica*

At the end of the month, each team receives a certificate of participation from IASC. A week later, their reports are evaluated by IASC for true or false asteroid reports. If they are true, they become preliminary discoveries. These will be further evaluated for six months to one year to determine if they are real asteroids. If they are, they become provisional discoveries and will be given provisional numbers and cataloged. After some years, the team will be able to name their discoveries. Additionally, this asteroid research citizen science project is part of NASA's planetary defense program, which aims to monitor asteroids that can potentially hit Earth in the future.

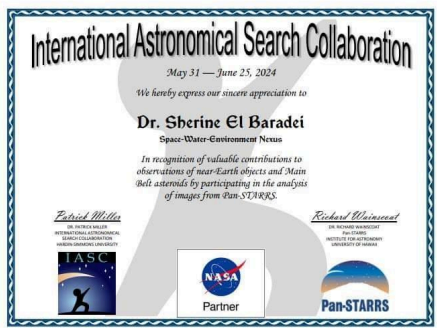
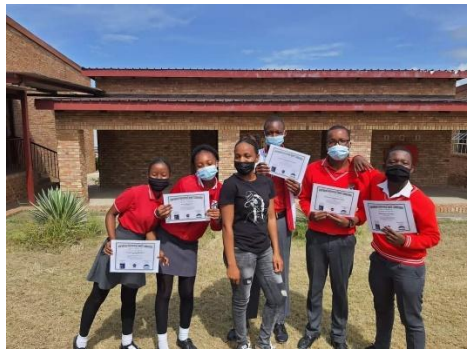
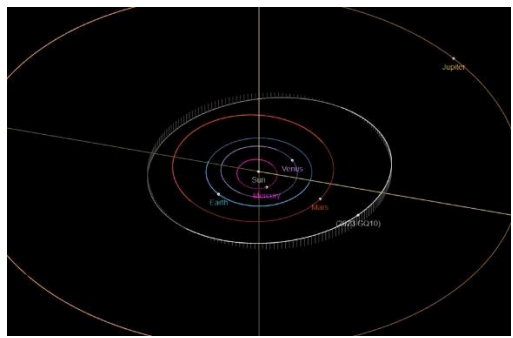

*Figure 6: The first picture shows the IASC certificates and the second picture is students posing with their IASC certificates of achievement after their research. The third picture is an aerial map of one of the many asteroids discovered by one of PACS e-Lab teams,*

2.4 Participant Involvement

When PACS e-Lab management sends out invitation letters, the primary focus is on entities with access to local community members, such as space education and outreach departments of national space agencies, astronomical observatories, STEM departments of African universities, astronomy organizations, secondary schools, and astronomy groups and amateurs. These entities typically have existing outreach programs, some resources to execute projects, and access to the public. PACS e-Lab communicates with the leaders of these groups through the general WhatsApp group and other means, providing training and sharing information, which they then disseminate to their members.

The asteroid search project is suitable for all individuals. The only requirements are having an adequate internet connection, a computer with a Windows operating system (Windows 7 and above), and a passion for astronomy. PACS e-Lab welcomes everyone in Africa who wishes to join, regardless of age, religion, country, sexual orientation, academic or professional affiliation and levels, etc., including both amateurs and professionals.

**2.5** Monitoring and Evaluation

To assess the impact of the asteroid research efforts over the years, a survey was distributed to the team leaders of each recruiting unit across the continent. They were asked to complete the survey by responding to the following questions: 1. the full name(s) of the team leadership. 2. The names of their countries and geographical zones. 3. The first year the group joined PACS e-Lab. 4. Name of the group/school/club/organization. 5. Age distribution of their participants. 6. Full names, gender identity, highest level of education, year of entry, and projects participated in, of all team members of that unit. 7. Comments on the experiences of team members towards the projects. 8. How the unit is adopting PACS e-Lab projects to the classroom, club, or community. 9. If the team has participated in any other projects outside PACS e-Lab. 10. Some of their challenges in participating or recruiting participants. 11. Their future plans regarding their participation and dissemination of the projects. Over 70 responses were received within a week, providing the data necessary to interpret the impact of PACS e-Lab. The results are presented in the section below.

Note: The survey data mentioned above were also utilized for a related research project titled "*Pan-African Citizen Science e-Lab: An Emerging Online Platform for Astronomy Research, Education and Outreach in Africa*" which is currently under review. Participants were informed that their responses might be used in multiple studies. The survey was designed to gather comprehensive feedback on the activities and impact of the PACS e-Lab initiatives, including the asteroid search project.

Also, the PACS e-lab requested the list of all asteroids discovered by all African citizen scientists from IASC, from where the management curated the names of those affiliated with the PACS e-lab. There were 52 enlisted discoveries but 32 are affiliated with PACS e-Lab and the latter is shown in the result section.

**3.0 Results**

Table 1 is about the geographical zones, the country, and the number of active asteroid citizen scientists associated with the team names, Table 2 shows the discovered asteroids with their provisional identities, the discoverers, their countries, and the date of discovery. The data from the survey were also used to prepare the charts below. Figure 9 is the picture of the African map showing the team's countries that have recorded success in the asteroid research and those that are still trying.

**Table 3.** The table below provides a list of geographical zones along with the number of active asteroid citizen scientists affiliated with each country and their team names.

| Zones | Country | Unit Names | Citizen Scientists |
|---|---|---|---|
| **North Africa** | Algeria | Algeria AstroResearcher | 2 |
| | Egypt | IEEE- A student breach activity | 28 |
| | Egypt | Pharaohs of Space | 2 |
| | Egypt | Space -Water-Environment Nexus e - Center | 44 |
| | Libya | Amateur Astronomy Libya | 7 |
| | Morocco | Steps Into Space Team (SIS) | 7 |
| | Morocco | Oukaimeden Observatory | 38 |
| | Tunisia | Tunisian Astronomical Society | 16 |
| | Tunisia | Aljarid Astronomie | 4 |
| **West Africa** | Benin | Sirius Astro-Club Benin | 6 |
| | Burkina Faso | Young Burkinabe Astrophysicists | 2 |
| | Côte d'Ivoire | Association Ivoirienne d'Astronomie | 39 |
| | Gambia | Physics Dept. University of Gambia | 8 |
| | Ghana | PRAGSAC | 4 |
| | Ghana | Ghana Planetarium | 6 |
| | Liberia | Center for Science Education | 4 |
| | Niger | Niger Space Surfer | 7 |
| | Nigeria | The Astro Group of the Rivers State University | 11 |
| | Nigeria | Phys. Dept. Chukwuemeka Odumegwu Ojukwu University Uli Campus Anambra Stat | 2 |
| | Nigeria | Astronomers Without Borders Nigeria/PACS e-Lab Nigeria. | 44 |
| | Nigeria | Arewa Astronomical Society | 1 |
| | Senegal | Orion Astrolab and Department of Physics, University Cheikh Anta Diop | 8 |
| | Togo | NGO Science Géologique pour un Développement Durable (SG2D | 5 |
| **Central Africa** | Angola | Associação Angolana de Astronomia | 5 |
| | Cameroon | Astronomy Club of the University of Buea | 9 |
| | CAR | IT Dreams and Promises of the University of Bamgui | 1 |
| | Congo | University Marien Ngouabi | 4 |
| | Congo | Denis Sassou N'GUESSO University | 9 |
| | DRC | Astroclub Kongo Central | 11 |
| | DRC | Lubumbashi Astro Club | 4 |
| | Gabon | NAEC - Gabon | 5 |
| **East Africa** | Burundi | Physics Dept. University of Burundi | 6 |
| | Ethiopia | Ethiopia Space Science Society Citizen Science Team | 23 |
| | Kenya | Space Partnerships and Applications Company Kenya | 8 |
| | Kenya | Kenya Space Agency - Education and Outreach Students' Network | 3 |
| | Kenya | Amateur Astronomical Society of Kenya | 14 |

|  | Mauritius | IAU - NOC Mauritius | 2 |
|---|---|---|---|
|  | Madagascar | Haikintana Astronomy Association | 23 |
|  | Rwanda | Physics Department, University of Rwanda, College of Science and Technology | 5 |
|  | Sudan | Sudanese Asteroid Hunters | 6 |
|  | Sudan | Institute of Space Research and Aerospace (ISRA) | 6 |
|  | South Sudan | Mayardit Academy for Space Sciences - University of Juba | 10 |
|  | Somalia | Somali Space Explorers | 1 |
|  | Tanzania | Department of Physics, University of Dodoma | 15 |
|  | Uganda | National Curriculum Development Centre of Uganda | 33 |
| **Southern Africa** | Zambia | Students from the Copperbelt University | 5 |
|  | Zambia | Zambia Space Explorers | 7 |
|  | Zambia | Kabulonga Girls' Secondary School | 1 |
|  | Mozambique | Detetives do Cosmos | 11 |
|  | Malawi | Celestial Explorers | 1 |
|  | Zimbabwe | Zimbabwe National Geospatial and Space Agency | 7 |
|  | Zimbabwe | Zimbabwe Astronomical Society | 8 |
|  | Botswana | University of Botswana Astronomy Club | 17 |
|  | Namibia | Department of Physics, Chemistry and Material Science University of Namibia | 16 |
|  | South Africa | BlueCraneSpace Astronomy & Astrophysics Department - University of Pretoria | 15 |
|  | South Africa | Night Sky Tours | 1 |
|  | South Africa | Galaxy Explorers | 5 |
|  | Lesotho | Maloti Space Explorers | 1 |

**Table 4:** lists the discovered asteroids along with their provisional identities, the names of the discoverers, their respective countries, and the dates of discovery.

| S/N | Provisional Asteroid ID | Citizen Scientists | Country | Date |
|---|---|---|---|---|
| 1 | 2021 OF15 | E. Chidozie, V. Akinola | Nigeria | July 31, 2021 |
| 2 | 2021 PV141 | S. Abotsi-Masters, R. Druguet | Ghana | August 5, 2021 |
| 3 | 2021 PR160 | O. Okolo | Nigeria/Ghana | August 5, 2021 |
| 4 | 2021 RZ190 | J. Olayiwola, A. Uchechukwu, O. | Nigeria/South Africa | September 4, 2021 |
| 5 | 2021 TK43 | N. Assamo, E. Jequecene, A. Izidine, H. Tandane | Mozambique | October 3, 2021 |
| 6 | 2021 VD60 | S.Aoko | Kenya | November 11, 2021 |
| 7 | 2021 VG27 | K. Gbaba, D. Addor, A. Anoukoum, E. Assih, A. Adjinare | Togo | November 1, 2021 |
| 8 | 2021 VU48 | K. Gbaba, D. Addor, A. Anoukoum, E. Assih, A. Adjinare | Togo | November 1, 2021 |
| 9 | 2022 BA34 | P.Pius, Lodrick, D.Mazengo, Y.Idala | Tanzania | January 28, 2022 |
| 10 | 2022 BA56 | P.Pius, Lodrick, D.Mazengo, Y.Idala | Tanzania | January 28, 2022 |
| 11 | 2022 BD31 | S. Yahaya, L. Yahaya, S. Yahaya | Niger | January 26, 2022 |
| 12 | 2022 CG36 | S. Yahaya, L. Yahaya, S. Yahaya | Niger | January 26, 2022 |
| 13 | 2022 BN25 | S. Yahaya, L. Ngounou, S. Yahaya, W. Yahaya, Y. Yahaya, N. Ngoulou | Niger | January 31, 2022 |
| 14 | 2022 DH10 | A. Idris | Nigeria | February 26, 2022 |
| 15 | 2022 DT10 | V.Chemane, L.Zita, G.Massunguine | Mozambique | February 27th, 2022 |
| 16 | 2022 GA12 | B.Mosinki, M. Kgobathe, R. Mokgethi, S. Rantsudu | Botswana | April 7, 2022 |
| 17 | 2022 GC13 | B.Mosinki, M. Kgobathe, R. Mokgethi, S. Rantsudu | Botswana | April 7, 2022 |
| 18 | 2022 HK11 | M. Rogers, C. Santus, N. Betty, B. Daniel, A. Sharon | Uganda | April 27, 2022 |
| 19 | 2022 OO43 | D. Yahaya, D. Yahaya, S. Yahaya, N. Ngoulou, Y. Yahaya, W. Yahaya | Niger | July 26, 2022 |
| 20 | 2022 SA192 | C. Ezeakunne | Nigeria | September 18, 2022 |
| 21 | 2022 SB170 | J. Mosinki, S. Rantsudu, M. Kgobathe | Botswana | September 26, 2022 |
| 22 | 2022 SD161 | T. Ngwane, M. Ncube, A. Nyamandi | Zimbabwe | September 19, 2022 |
| 23 | 2022 SU184 | T. Ngwane, M. Ncube, A. Nyamandi | Zimbabwe | September 19, 2022 |
| 24 | 2022 SK100 | A. Thomas | Liberia | September 19, 2022 |
| 25 | 2022 SO221 | A. Thomas | Liberia | September 19, 2022 |
| 26 | 2022 SR255 | A. Thomas | Liberia | September 19, 2022 |
| 27 | 2022 SX223 | A. Thomas | Liberia | September 19, 2022 |
| 28 | 2022 SZ156 | A. Thomas | Liberia | September 19, 2022 |
| 29 | 2022 SL97 | M. Mohamed | Sudan | September 22, 2022 |
| 30 | 2022 SW188 | A. Diaby, S. Ahoua | Ivory Coast | September 18, 2022 |
| 31 | 2022 UJ79 | A. Rajaonarivelo, M. Andriatiana, V. Randriamanantena, T. Rabarison, M. Rabeony, A. Rakotomandimby, A. Rabarisoa, N. Ramanitrandrasana, H. Rabekoto, T. Rabarison | Madagascar | October 19, 2022 |
| 32 | 2023 GQ10 | Y. Wyk, A. Ndlovu, N. Wright | South Africa | April 15, 2023 |

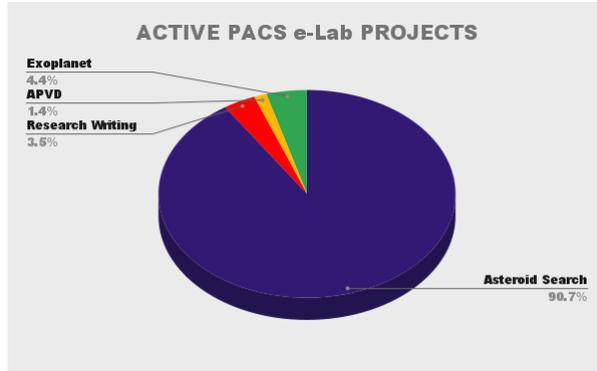

*Figure 7*: Pie chart illustrating that the asteroid search endeavor is the most significant among the four PACS e-Lab projects.

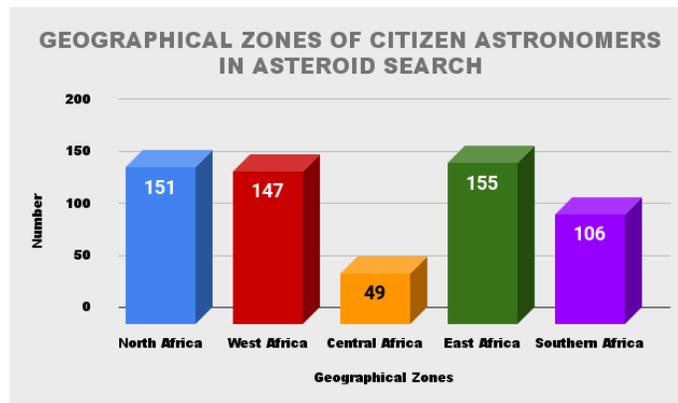

*Figure 8*: Bar chart comparing the numbers of active asteroid citizen scientists across the five geographical zones in Africa.

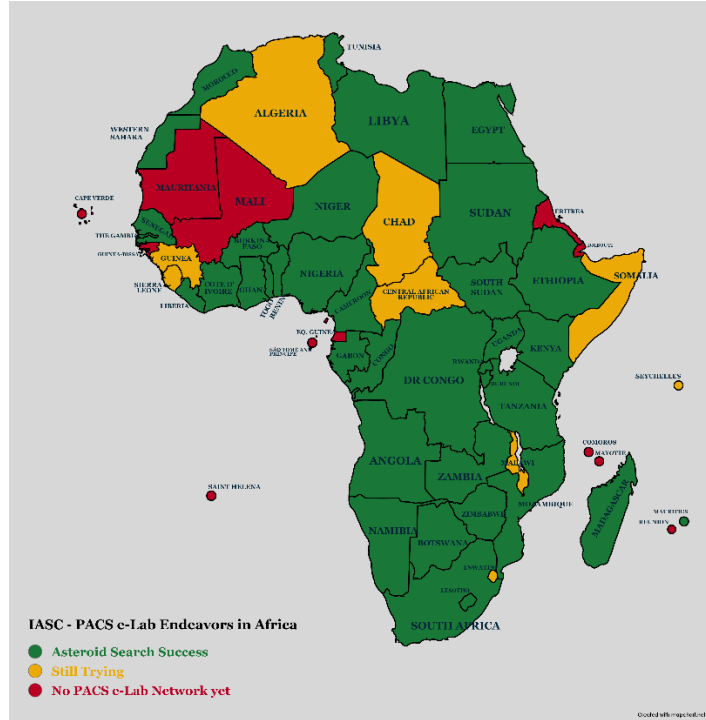

*Figure 9*: Map of Africa highlighting the countries where teams have achieved success in asteroid research and those that are still working towards it.

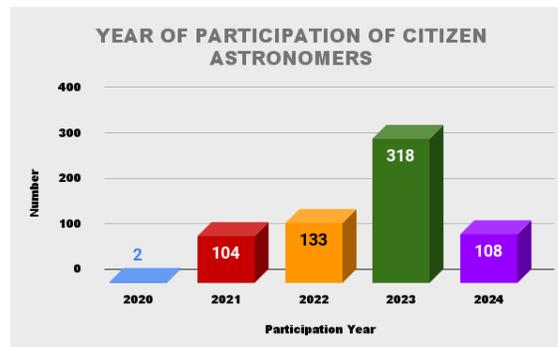

Figure 10: Chart showing the number of citizen astronomers' recruitment each year.

## 4.0 Discussion

The summary of the results is given in the table 5 below

*Table 5: Summary of data from Tables 3 and 4.*

| Groups/Units/Teams | Countries | Total number of active Asteroid Citizen Scientists | Provisional Discoveries | No. of Citizen Scientists with Provisional Asteroid Discoveries. |
|---|---|---|---|---|
|  |  |  |  |  |

| 58 | 40 | 595 | 32 | ≥ 60 |

From the survey, the total number of citizen astronomers involved in the asteroid search is 595, coming from 58 teams in 40 countries across the continent

Tables 4 and 5 show that a total of 32 asteroids have been discovered by over 60 of the citizen scientists, most of whom worked in teams, with a few working alone. Mr. Allen Thomas, a secondary school principal from Liberia in 2022, recorded the highest number of asteroid discoveries by an individual, with a total of 5 discoveries.

At PACS e-Lab, the asteroid search is the leading project with 90.7% participation as shown in Figure 7. In terms of geographical participation in Figure 8, East Africa leads, followed by North Africa, West Africa, Southern Africa, and lastly Central Africa.

Figure 9 shows the asteroid research endeavors in Africa by country. The ones colored in green have achieved success in the asteroid search, meaning they have mastered the research from the entry knowledge levels to advanced ones. Most of them have more than one independent group, and they are very active and always recruiting and training more people. The yellow-colored ones are still working towards becoming like the green, and there is only one team from those countries working with PACS e-Lab. The red countries are places where PACS e-Lab has not had any presence yet and the management is improving its methodology to recruit teams from there.

Tables 1 and 2 are samples of a typical academic calendar issued by IASC to PACS e-Lab before the beginning of the research each year.

**5.0 Conclusion**

This paper presents the asteroid search activities in Africa through the collaborative efforts of PACS e-Lab and IASC. The research aims to foster astronomy research, education, and outreach on the continent while also enabling Africa to contribute to global planetary defense efforts against asteroids and other space debris.

Since December 4, 2020, until the time of writing in July 2024, PACS e-Lab has expanded the project across more than 40 African countries, benefiting nearly 600 people. PACS e-Lab plans to continue recruiting individuals from both active countries (Green and Yellow countries) and those currently inactive (Red countries) as shown in Figure 9. The management aims to recruit and train thousands of citizen scientists from all 54 African countries in asteroid search projects.

Tables 4 and 5 show that over 60 out of the 595 citizen scientists have discovered a total of 32 asteroids, with most working in teams and a few working alone. In 2022, Mr. Allen Thomas, a secondary school principal from Liberia, recorded the highest number of asteroid discoveries by an individual, with a total of 5 discoveries.

The number of provisional asteroid discoveries recorded from the list sent by IASC is 52, with 32 of them discovered by those affiliated with PACS e-Lab. This indicates that while other groups are participating independently in the research, PACS e-Lab's efforts lead and dominate the continent. Moreover, PACS e-Lab had its highest number of recruits in 2023 as shown in Figure 10, with many individuals recording tens of preliminary asteroid discoveries. Many of these may become provisional discoveries, so the current number of 32 is merely temporary. PACS e-Lab management projects that this number might reach 60 if all these preliminary discoveries are confirmed.

Asteroid search is the most prominent of all PACS e-Lab projects. It was the first project at PACS e-Lab for over two years before others were incorporated. Even after the incorporation of other projects, asteroid search remained a prerequisite for participation in all projects. However, this rule is no longer compulsory.

Tables 1 and 2 show that the research is ongoing every month. During the 2024/2025 academic year, IASC has granted PACS e-Lab 35 slots, meaning about 35 groups can participate in the research each month. In light of this, PACS e-Lab is extending its partnerships to more organizations across the continent to join forces in engaging the African public in research.

## Data Availability Statement

All data used for this study were obtained through a survey. The survey data were summarized in a Google Sheet, which is included in this article. The original survey data will be made available upon request.

## Acknowledgment

The PACS e-Lab teams are grateful to their international partners: International Astronomical Search Collaboration (IASC), for their partnership

PACS e-Lab management also thanks Dr. Charles Takalani, the head of the Secretariat for the African Astronomical Society, for always being there for PACS e-Lab, assisting in various ways such as providing supporting letters and signatures for PACS e-Lab's international collaboration whenever needed

PACS e-Lab management is grateful to Ezeakunne Chidozie (PhD candidate) for donating to PACS e-Lab to support its efforts.

PACS e-Lab extends warm appreciation to all its volunteers and team leadership for their efforts.

## Conflict of interest

The authors declare that there are no potential conflicts of interest in this publication.

## Ethical statement

All the participants were informed about the publication and they granted their consent to the work.

## Authors' Contribution

All authors have been involved in the asteroid search project. Miracle Chibuzor Marcel wrote the drafts of the paper. The rest of the authors, who are team leaders and members of various units, filled out the survey that provided data for this study.